# Thermal Contact Resistance Across Nanoscale Silicon Dioxide and Silicon Interface

Jie Chen,[1] Gang Zhang,[2, *] and Baowen Li[1, 3, 4]

[1]Department of Physics, Centre for Computational Science and Engineering, and Graphene Research Centre, National University of Singapore, Singapore 117542, Singapore

[2] Key Laboratory for the Physics and Chemistry of Nanodevices and Department of Electronics, Peking University, Beijing 100871, People's Republic of China

[3]NUS Graduate School for Integrative Sciences and Engineering, Singapore 117456, Singapore

[4]NUS-Tongji Center for Phononics and Thermal Energy Science and Department of Physics, Tongji University, Shanghai 200092, People's Republic of China

*E-mail: zhanggang@pku.edu.cn

## Abstract

Silicon dioxide and silicon ($SiO_2$/Si) interface plays a very important role in semiconductor industry. However, at nanoscale, its interfacial thermal properties haven't been well understood so far. In this paper, we systematically study the interfacial thermal resistance (Kapitza resistance) of a heterojunction composed of amorphous silicon dioxide and crystalline silicon by using molecular dynamics simulations. Numerical results have shown that Kapitza resistance at $SiO_2$/Si interface depends on the interfacial coupling strength remarkably. In the weak interfacial coupling limit, Kapitza resistance depends on both the detailed interfacial structure and the length of the heterojunction, showing large fluctuation among different samples. In contrast, it is almost insensitive to the detailed interfacial structure or the length of the heterojunction in the strong interfacial coupling limit, giving rise to a nearly constant value around $0.9\times10^{-9}$ $m^2KW^{-1}$ at room temperature. Moreover, the temperature dependent Kapitza resistance in the strong interfacial coupling limit has also been examined. Our study provides useful guidance to the thermal management and heat dissipation across nanoscale $SiO_2$/Si interface, in particular for the design of silicon nanowire based nano electronics and photonics devices.

# 1. INTRODUCTION

The interface between silicon dioxide and silicon ($SiO_2$/Si) is the basis for most current Si-based microelectronics technology. In recent years, silicon nanowires have attracted much attention and have shown promising applications as the building blocks for nanoelectronic devices [1-3]. Experimental study has reported the growth of straight silicon nanowires on $SiO_2$ substrate with uniform diameter, length and orientation [4], which are important factors for the practical applications of silicon nanowire based nanoelectronic devices. Nanoelectronic devices can generate huge heat flux in very small areas (also known as hot-spot). As the silicon stacked chips or three-dimensional chips are usually investigated, this can create smaller and hotter spots. Hot-spot removal is a key for the future generation integrated nanoelectronics. The thermal contact/interfacial resistance plays a critical role in the transport of thermal energy in nano devices. Therefore, a complete understanding of nanoscale interfacial thermal transport properties is vital in the integration of nano devices, in particular for the silicon nanowire based electronics, photonics and energy conversion applications.

The nanoscale thermal contact resistance of highly perfect interfaces in epitaxial TiN and carbon nanotube systems has been experimentally measured [5,6]. But the interpretation of nanoscale amorphous $SiO_2$ and Si interface remains controversial, because of the complexities inherent in studying disordered materials. Compared to the intensive study on the electronic properties [7-9] of $SiO_2$/Si interface, its interfacial thermal properties are much less explored.

The commonly used theoretical models to predict the thermal contact resistance include the acoustic mismatch model (AMM) and diffusive mismatch model (DMM) [10]. The AMM assumes specular scattering at the interface, and the phonon transmission and reflection are calculated from the mass density and anisotropic elastic constants of materials. The diffusive mismatch model assumes that phonons are randomly and elastically scattered at the interface, and the transmission coefficient is determined by the ratio of the densities of vibrational states on either side of the interface. Although these models have greatly advanced the understanding of thermal transport across interface, they predict thermal contact resistance based on assumptions about the nature of phonon scattering at the interface. In practical situations, the degree of specular and diffusive scattering depends on the quality of the interface [10], and can only be modeled qualitatively [11]. In AMM and DMM models, the phonon dispersion relation is usually approximated by a linear dispersion [12], which is accurate for wave vectors close to the zone center, but deviates significantly for wave vectors near the zone edges. Moreover, the DMM model describes only a singular diffusive scattering process. Therefore, it underestimates the thermal contact resistance in some cases [5, 13, 14], while overestimates the thermal contact resistance in other case [14]. More importantly, the atomic level details of the interfacial structures are neglected in both models, which can lead to inaccurate prediction of thermal contact resistance at temperature where phonons with wave length on the same scale as the interatomic spacing are excited [15]. Thus the atomistic level approach is indispensable, and has been widely used to study the interfacial thermal properties in various material systems [15-20].

In this paper, by using silicon dioxide and silicon nanowire junctions as examples, we systematically study the interfacial thermal resistance (Kapitza resistance) at nanoscale $SiO_2$/Si interface by using molecular dynamics (MD) simulations, which is an atomic level approach and has no assumption about the nature of the phonon scattering at the interface. Numerical results have shown that the Kapitza resistance at SiO$_2$/Si interface depends on the interfacial coupling strength remarkably and shows distinct dependence on system parameters with different coupling strength. Moreover, the temperature dependent Kapitza resistance in the strong interfacial coupling limit has also been discussed. Our study provides useful guidance to the thermal management and heat dissipation in silicon-based nano devices.

## 2. RESULTS AND DISCUSSION

Our modeling system is a heterojunction composed of amorphous silicon dioxide (*a*-SiO$_2$) and crystalline silicon (*c*-Si). The interatomic forces between *c*-Si atoms are calculated according to Tersoff potential [21], which has been widely used to study the lattice dynamics [22], thermal and structural properties [23], thermomechanical properties [24], point defects [25], and the liquid and amorphous phases of Si [23, 26]. To describe atomic interaction in *a*-SiO$_2$, a modified parameter set for Tersoff potential based on *ab initio* calculations is used [27].

Recent study on the interfacial thermal resistance (ITR) at the solid-solid interface has shown that ITR depends only on the dimensions along the direction of heat conduction, rather than those perpendicular to it [17]. Therefore, the MD simulation domain in our study has a fixed cross section area of 17×17 Å$^2$ and adjustable length in the longitudinal direction. Here we set longitudinal direction along *x*-axis. Velocity Verlet algorithm is employed to integrate Newton's equations of motion, and each MD step is set as 0.5 fs. For *c*-Si segment, its atomic structure is constructed from diamond structured bulk silicon, with *x*-axis oriented along the [100] direction. To generate *a*-SiO$_2$ segment at a given temperature $T_0$, we start with the crystalline form of silicon dioxide: alpha-quartz (*α*-quartz). We apply Langevin heat bath to equilibrate *α*-quartz at 3000 K (above melting point) for 100 ps in order to achieve the amorphous structure. The resultant structure is then annealed to temperature $T_0$ with a constant cooling rate of $10^{13}$ K/s [27]. This approach to generate amorphous $SiO_2$ has been used in the construction of $c$-Si/$a$-$SiO_2$ core/shell nanowires [28].

After annealing, $c$-Si segment with the same cross section area is coupled to $a$-$SiO_2$ segment with an adjustable separation distance $L_s$, which is defined as the minimum separation along $x$-axis between two segments at the interface (Fig. 1). In this way, the strength of interfacial coupling in our modeling is controlled by the separation distance $L_s=L_0/N_c$, where $L_0=5.43$ Å is the lattice constant of $c$-Si and $N_c$ is an adjustable parameter. In our simulation, the length of two segments are set equal and the total length of the heterojuction is $L_x$. In the previous experimental study on the chemical structure of $SiO_2$/Si interface [29], it was found that there is a transition region of altered structure between the crystalline silicon and the amorphous silicon oxide. This transition region is considered to be formed by stress between the two segments. Our interface model is generally consistent with the idea of transition region. The heterojunction is then attached to Langevin heat bath at temperature $T_0$ for 100 ps to relax the structure and reach thermal equilibrium. During this relaxation process, the net angular momentum is removed at each step [30] to avoid the torsion at the interface. Moreover, the neighbor list is dynamically updated every ten time steps, and all atoms are allowed to move freely in all directions according to the interatomic interaction.

Fig. 2 shows the detailed interfacial structure for different $SiO_2$/Si samples after structure relaxation. Since the maximum cut-off distance for different chemical bonds is 3 Å in Tersoff potential [21, 27], $N_c=2$ corresponds to the weak interfacial coupling case. As shown in Fig. 2(a-b), two segments are weakly connected by a few interfacial bonds in the weak coupling case ($N_c=2$), giving rise to a sharp interface. Moreover, the number of the connected bonds depends on the detailed interfacial structure of $a$-SiO$_2$ segment, and can vary significantly among different samples. In contrast, two segments are densely connected by many interfacial bonds in the strong coupling case ($N_c$=20) with small fluctuation among different samples (Fig. 2(c-d)). In this case, the interface is less obvious compared to the weak coupling case.

After structure relaxation, we apply nonequilibrium MD simulation to calculate the temperature profile and heat flux along the heterojunction. In the longitudinal direction, fixed boundary condition is imposed at the two ends. Free boundary condition is used to surface atoms. Next to the boundary layers at both ends, Langevin heat baths with different temperature are applied as the heat source and sink. The temperature of two heat baths is set as $T_H=T_0+\Delta/2$ and $T_L=T_0-\Delta/2$, respectively, where $T_0$ is the mean temperature and $\Delta$ is the temperature difference. In all simulations, we keep the temperature difference small ($\Delta/T_0$<7%). The heat flux $J$ is defined as the energy injected into/removed from the heat source/sink across unit area per unit time [31]. In a nonequilibrium steady state, these two rates are equal. The local temperature along $x$-axis in the neighborhood of location $x$ is defined as

$$T(x) = \left\langle \sum_{i=1}^{N} m_i \vec{v}_i \cdot \vec{v}_i \right\rangle / (3Nk_B), \qquad (1)$$

where $N$ is the total number of atoms within the neighborhood, $k_B$ is the Boltzmann constant, $m_i$ and $\vec{v}_i$ is the mass and velocity of atom $i$, respectively, and the bracket denotes the temporal average. Our simulations are performed long enough (~10$^7$ time steps) to allow system to reach the steady state in which both the temperature profile and heat flux are almost constant. Finally, Kapitza resistance is calculated according to

$$R = \Delta T / J, \qquad (2)$$

where $\Delta T$ is the temperature drop at the interface, and $J$ is the heat flux across the interface. In order to minimize the uncertainty in determining $\Delta T$, a least-square linear regression analysis described by Landry and McGaughey [15] has been applied to the temperature profile. With this analysis, the temperature drop at the interface can be determined by extrapolating the linear fit lines of two segments to the interface ($x/L_x$=0.5) and then calculating the temperature difference. For a given system parameter (e.g., coupling strength, length, temperature), the final calculation results are averaged over eight samples with different interfacial structures.

Fig. 3 shows the temperature profile for different $SiO_2$/Si samples with the length about $L_x$=19.2 nm at room temperature. With weak interfacial coupling, there exists an obvious temperature drop at the interface (Fig. 3(a)). Moreover, this temperature drop can vary significantly among different samples when the interfacial coupling is weak. For instance, for the two samples shown in Fig. 3(a), the calculated temperature drop is 9.0 K (square) and 4.9 K (circle), and the corresponding ITR in these two samples are $3.84\times10^{-9}$ $m^2KW^{-1}$ and $1.71\times10^{-9}$ $m^2KW^{-1}$, respectively. In this case, ITR in one sample is almost twice as large as that in another sample. These results suggest that in the weak interfacial coupling limit, ITR of $SiO_2$-Si depends sensitively on the detailed interfacial structure, such as number of connected bonds, type of connected bonds, and angle/distance between connected bonds. As demonstrated in Fig. 2(a-b), these factors can vary significantly among different samples in the weak coupling case, leading to the large fluctuation in ITR. In contrast, the temperature drop in the strong interfacial coupling case is much smaller and less obvious compared to the weak coupling case (Fig. 3(b)), giving rise to ITR values with small fluctuation among different samples. For instance, for the two samples show in Fig. 3(b), the corresponding ITR are $0.93\times10^{-9}$ $m^2KW^{-1}$ and $1.14\times10^{-9}$ $m^2KW^{-1}$, respectively. This much smaller fluctuation of ITR in the strong coupling case is because two segments are densely connected by a lot of interfacial bonds, and the difference in the detailed interfacial structure among different samples is insignificant (Fig. 2(c-d)).

To systematically study ITR at $SiO_2$/Si interface, we first consider the effect of interfacial coupling strength on ITR with fixed length of heterojunction. Fig. 4 shows ensemble averaged ITR at room temperature versus separation distance ($L_s$) between two segments (each segment has a fixed length of 2.8 nm). In the weak coupling limit ($N_c$=2, $L_s$=2.7 Å), the ensemble averaged calculation result of ITR is (4.27±1.78)×$10^{-9}$ $m^2KW^{-1}$. This value of ITR is higher than that of Si/Ge interface (~3.0×$10^{-9}$ $m^2KW^{-1}$) [15, 17], graphene/phenolic resin interface (~0.5×$10^{-9}$ $m^2KW^{-1}$) [18], and graphene/graphane interface (~$10^{-11}$ $m^2KW^{-1}$) [19] at the same temperature. With the decrease of $L_s$, the coupling strength between two segments becomes stronger, revealed by the densely connected interfacial bonds (Fig. 2(c-d)). In the weak coupling case, both analytical and numerical investigations have shown that the heat flux across the interface is proportional to the square of interfacial coupling strength [32, 33], so that ITR at the weakly coupled interface can be reduced by increasing interfacial coupling strength [34]. As a result, ITR decreases monotonically with the decrease of $L_s$, showing a nearly exponential dependence on $L_s$. This exponential dependence of ITR on separation distance is presumably due to the exponentially short-range interaction in Tersoff potential [21]. The converged value of ITR is (0.83±0.25)×$10^{-9}$ $m^2KW^{-1}$ in the strong coupling limit ($N_c$=20, $L_s$=0.27 Å). Notice the standard deviation of ITR among different samples in the strong coupling limit is much smaller than that in the weak coupling limit, consistent with the interfacial structure shown in Fig. 2 and the temperature drop shown in Fig. 3. Our results are consistent with a recent study on ITR between two Lennard-Jones fcc solids which reports the increase of ITR at a given temperature when the strength of cross-species interactions decreases [35].

The difference in ITR values with different coupling strength can be qualitatively understood from the power spectrum of interfacial atoms. To calculate the power spectrum, we first equilibrate the $SiO_2$/Si sample with Langevin heat bath for 100 ps. After structure relaxation, nonequilibrium MD simulation runs for $5\times10^5$ time steps and the velocity of the interfacial atom is recorded at each step. Then the velocity auto-correlation function is calculated based on the recorded velocity. Finally, the power spectrum is calculated from the Fourier transform of atom's velocity auto-correlation function. To ensure the generality of the power spectrum at the interface, we have checked explicitly the power spectrum for different atoms at the interface. We found the power spectrum is similar for different interfacial atoms on the same side of the interface.

Fig. 5 shows the typical power spectrum for atoms at each side of the $SiO_2$/Si interface with different interfacial coupling strength. With weak interfacial coupling, the power spectrum for interfacial atoms at two sides is almost decoupled from each other, showing the signature of phonon density of states for individual material at each side. For the $SiO_2$ side, the cut-off frequency is around 40 THz, and the glass peaks around 12 THz, 20 THz, and the high frequency doublet around 30 THz are well reproduced in our calculations (solid line in Fig. 5(a)), which are in good agreement with experimental data [36]. For the Si side, a typical phonon spectrum for Si is recovered (dashed line in Fig. 5(a)), with notable peaks at around 5 THz and 15 THz, consistent with previously reported phonon density of states for Si in literature [37, 38]. Due to the weak coupling, the power spectrum at each side is slightly mixed with each other. For instance, the power spectrum for the Si side is extended to 25 THz, compared to the cut-off frequency about 16 THz for pure Si [37, 38]. Moreover, additional high frequency peak around 20 THz shows up in the power spectrum for Si, while notable low frequency peaks (less than 10 THz) are introduced into the power spectrum for $SiO_2$. Due to the different cut-off frequency for Si and $SiO_2$, it is clearly shown in Fig. 5(a) that there exists an obvious mismatch in power spectrum above 25 THz in the weak coupling case. This large mismatch in power spectrum hinders the heat transport across the interface and thus leads to the large ITR value [34]. With the increase of interfacial coupling strength, the mix of power spectrum becomes more significant, leading to the further extension of cut-off frequency to 40 THz for the Si side (Fig. 5(b)). Due to the well mixed power spectrum, the ITR values in the strong coupling case are smaller than that in the weak coupling case.

Another factor that can affect the heat transport across the heterojunction composed of two dissimilar segments is the length of the heterojunction [17, 33]. In our study, we set the length of each segment equal to $L$ and increase the length of each segment simultaneously to investigate the length effect on ITR. Fig. 6 shows the room temperature ITR versus the length of the heterojunction $L_x$ ($L_x \approx 2L$) for both weak and strong interfacial coupling cases. In the weak coupling case ($N_c$=2), ITR decreases monotonically with the increase of $L_x$, from $(4.27\pm1.78)\times10^{-9}$ $m^2KW^{-1}$ at $L_x$=5.6 nm to $(1.93\pm1.04) \times10^{-9}$ $m^2KW^{-1}$ at $L_x$=76.8 nm, following a power law decay (solid line in Fig. 6). These results reveal that in addition to the detailed interfacial structure, the length of each segment can also affect the ITR quite significantly in the weak coupling case. The power law dependence of ITR on length is in line with recent study on ITR of Si/Ge interface [17]. In contrast, however, the segment length has little effect on ITR in the strong coupling case ($N_c$=20), giving rise to a nearly constant value of ITR about $0.9\times10^{-9}$ $m^2KW^{-1}$ with small fluctuation (circle in Fig. 6). Moreover, in the long length limit, the predictions of ITR in the weak and strong coupling cases partially overlap with each other. This result suggests that when the length of each segment is long enough, the interfacial properties of the heterojunction, such as the interfacial coupling strength and detailed interfacial structure, are no longer the dominant factor in determining ITR, consistent with previous theoretical study on the heat conduction in dissimilar nonlinear lattice models [33].

For the application of heat dissipation, the device usually works at high temperature. Therefore, we finally study the temperature dependent Kapitza resistance at $SiO_2$/Si interface above room temperature. As forementioned, ITR depends sensitively on the interfacial structure in the weak coupling limit. To generate $SiO_2$/Si interface, there are some random processes involved in our simulation, such as thermal equilibration and annealing. As a result, we cannot have the precise control of the detailed interfacial structure. In order to solely examine the temperature effect, we consider the strong coupling case in which ITR is much more robust with detailed interfacial structure compared to the weak coupling case. Fig. 7 shows the temperature dependent ITR for the strong coupling case ($N_c$=10). Above room temperature, ITR decreases monotonically with the increase of temperature. This is because with the increase of temperature, anharmonicity of the atomic interaction increases. The phonon transmission coefficient is enhanced through inelastic scattering and therefore decreases the thermal contact resistance [10]. Similar trends of temperature dependence of ITR have been reported in the studies of various interface systems both theoretically [15, 34, 39] and experimentally [5, 40, 41]. Our results demonstrate that inelastic phonon scattering is significant at $SiO_2$/Si interface above room temperature.

It is worth to compare the present MD results with previous related literatures. The prediction of ITR at 500 K in our study is (0.68±0.30)×$10^{-9}$ $m^2KW^{-1}$, which is in close agreement with a recent theoretical study on ITR of $a$-SiO$_2$/$c$-Si interface at the same temperature using a different approach [45]. In Ref. [45], Lampin *et al.* developed an alternative approach to extract ITR from the relaxation times of the heat flux exchanged across the interface. The reported ITR of $a$-SiO$_2$/$c$-Si interface in their study is $0.4\times10^{-9}$ m$^2$KW$^{-1}$ at 500 K. Moreover, thermal rectification phenomenon [46] has been observed in asymmetric nanoscale junctions, such as mass-graded nanotubes [47] and water/silica interfaces [48]. However, as amorphous silicon oxide is used in the present work, the number of interfacial bonds changes case by case, which leads to the fluctuation in calculated heat flux under certain temperature difference. And due to the small mass difference between the two ends, the rectification is ignored in this structure.

The results of nanoscale thermal contact resistance at interface between silicon and amorphous silicon oxide are helpful to understand the dramatically low thermal conductivity observed experimentally in silicon nanowires [42-44]. In experiment, there may exist amorphous silicon oxide between silicon nanowires and the metallic electrode. Thus the measured thermal resistance is the sum of the thermal resistance of pristine silicon nanowires and the thermal contact resistance. In principle, the presence of thermal contact resistance will lead to the experimentally measured thermal conductivity lower than the intrinsic thermal conductivity. However, using the experimentally measured room temperature thermal conductivity of silicon nanowires with diameter of 37 nm and length of 5 $\mu$m ($L$=5 $\mu$m) as example [44] (thermal conductivity $\kappa\approx$19 Wm$^{-1}$K$^{-1}$), we can find that even in the weak coupling condition ($R\sim4\times10^{-9}$ m$^2$KW$^{-1}$), including thermal contact resistance gives rise to the effective thermal conductivity about 18.4 Wm$^{-1}$K$^{-1}$ [49], only resulting in about 3% reduction of thermal conductivity. Thus the Kapitza resistance at SiO$_2$/Si interface has very weak influence on the experimentally measured thermal conductivity of silicon nanowires, due to the low thermal conductivity of pristine silicon nanowires. Our analysis suggests that the thermal contact resistance should not be the source of the ultralow thermal conductivity of silicon nanowire systems.

## 3. CONCLUSION

In summary, by using silicon dioxide and silicon nanowire junctions as examples, we have systematically studied Kapitza resistance at $SiO_2$/Si interface by using atomic level simulations. It is found that Kapitza resistance at $SiO_2$/Si interface decreases monotonically with the increase of interfacial coupling strength. In the weak interfacial coupling limit, Kapitza resistance depends remarkably on the detailed interfacial structure, leading to a large fluctuation of prediction results among different samples. Moreover, it shows a power law decay when the length of the heterojunciton increases. In contrast, in the strong coupling limit, Kapitza resistance is almost insensitive to the detailed interfacial structure or the length of the heterojunction, giving rise to a nearly constant value around $0.9 \times 10^{-9}$ $m^2KW^{-1}$ at room temperature. Furthermore, with the increase of temperature, Kapitza resistance in the strong interfacial coupling limit decreases monotonically above room temperature. Our study provides useful guidance to the thermal management and heat dissipation in silicon-based nano electronics and photonics devices.

**Acknowledgements**

This work was supported in part by a grant from the Asian Office of Aerospace R&D of the US Air Force (AOARD-114018), and a grant from SERC of A*STAR (R-144-000-280-305), Singapore. G.Z. was supported by the Ministry of Science and Technology of China (Grant No. 2011CB933001) and Ministry of Education, China (Grant No. 20110001120133). J.C. acknowledges the World Future Foundation (WFF) for awarding him the WFF PhD Prize in Environmental and Sustainability Research (2012) and the financial support to this work.

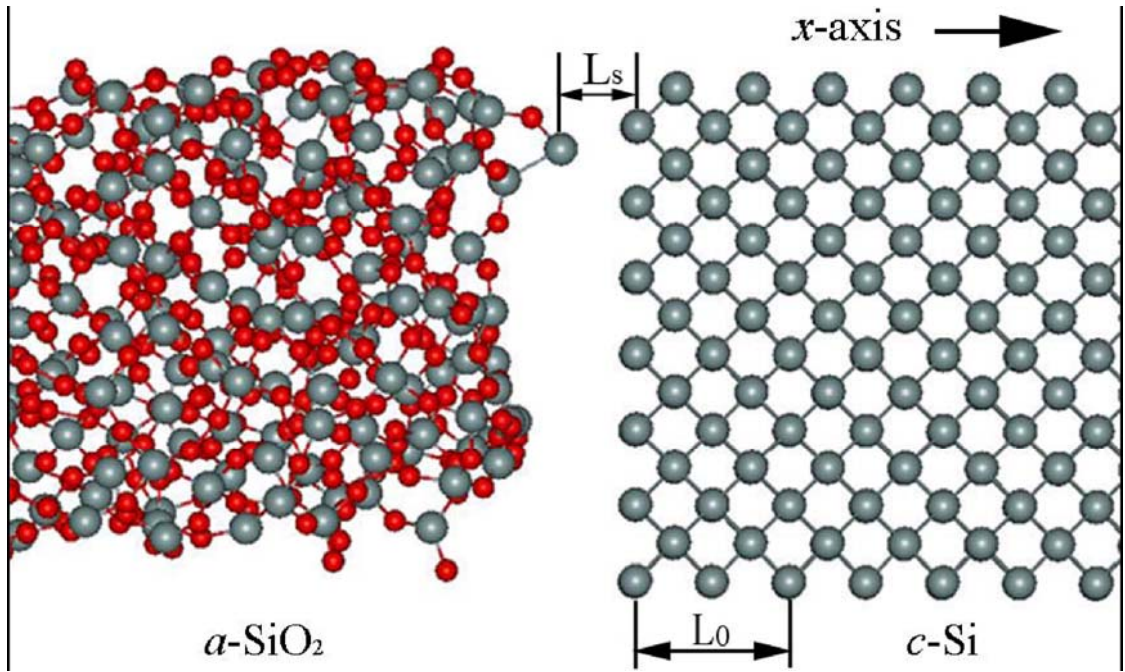

Figure 1. Side view of SiO$_2$/Si interface before structure relaxation at room temperature. The grey and red color denotes Si and O atom, respectively. The longitudinal direction is set along $x$-axis. After annealing, $a$-SiO$_2$ is coupled to $c$-Si with a minimum separation of $L_s=L_0/N_c$ along $x$-axis, where $L_0$ denotes the lattice constant of $c$-Si and $N_c$ is an adjustable parameter.

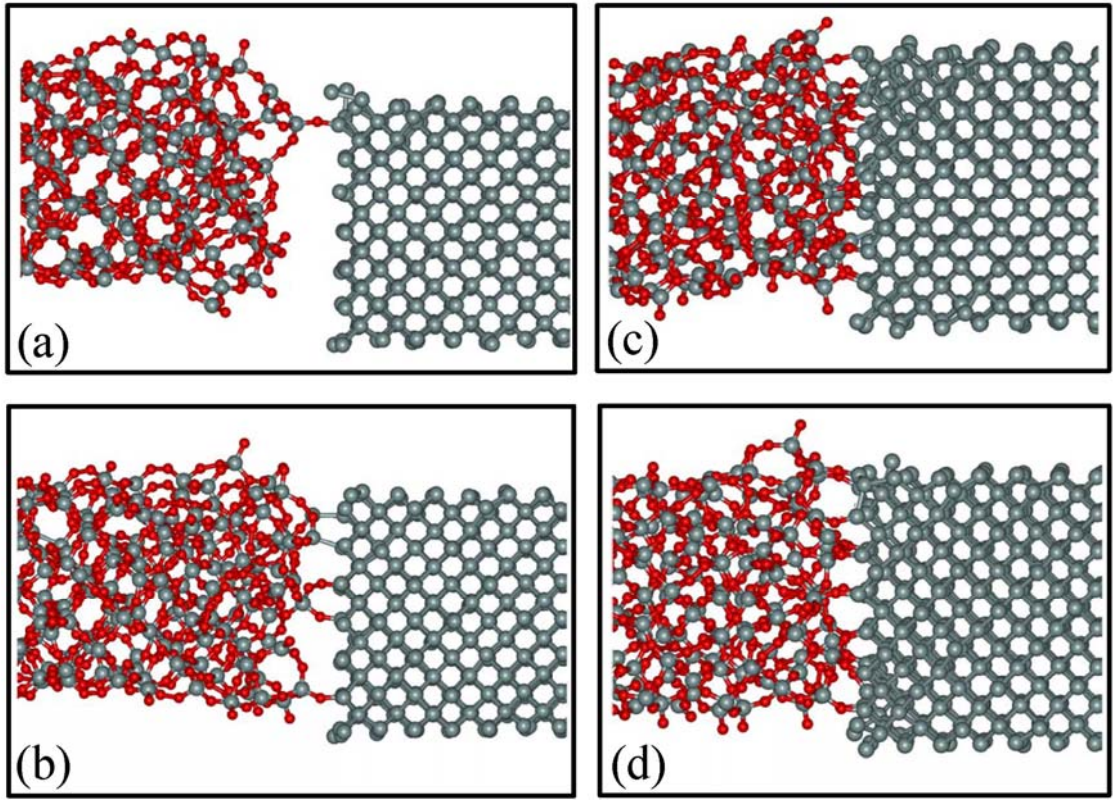

Figure 2. Side view of different $SiO_2$/Si samples after structure relaxation in the weak and strong interfacial coupling cases at room temperature. The grey and red color denotes Si and O atom, respectively. (a-b) Weak coupling case $N_c$=2. (c-d) Strong coupling case $N_c$=20.

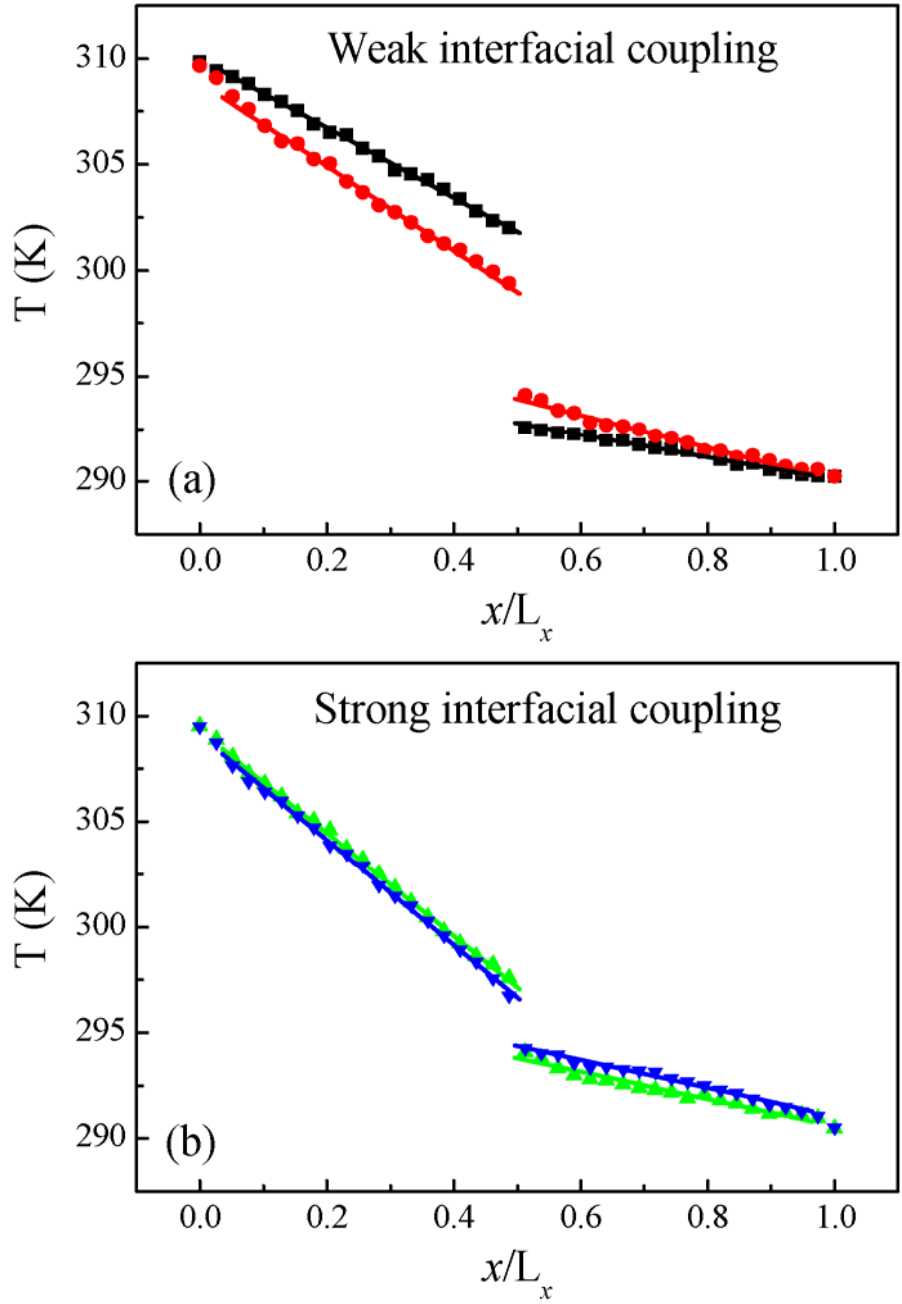

Figure 3. Temperature profile in the longitudinal direction for different $SiO_2$/Si samples at room temperature. The length of the heterojunction is about $L_x$=19.2 nm. Symbols denote simulation results for different samples, and solid lines denote the corresponding least-square linear fit. (a) Weak interfacial coupling ($N_c$=2). (b) Strong interfacial coupling ($N_c$=20).

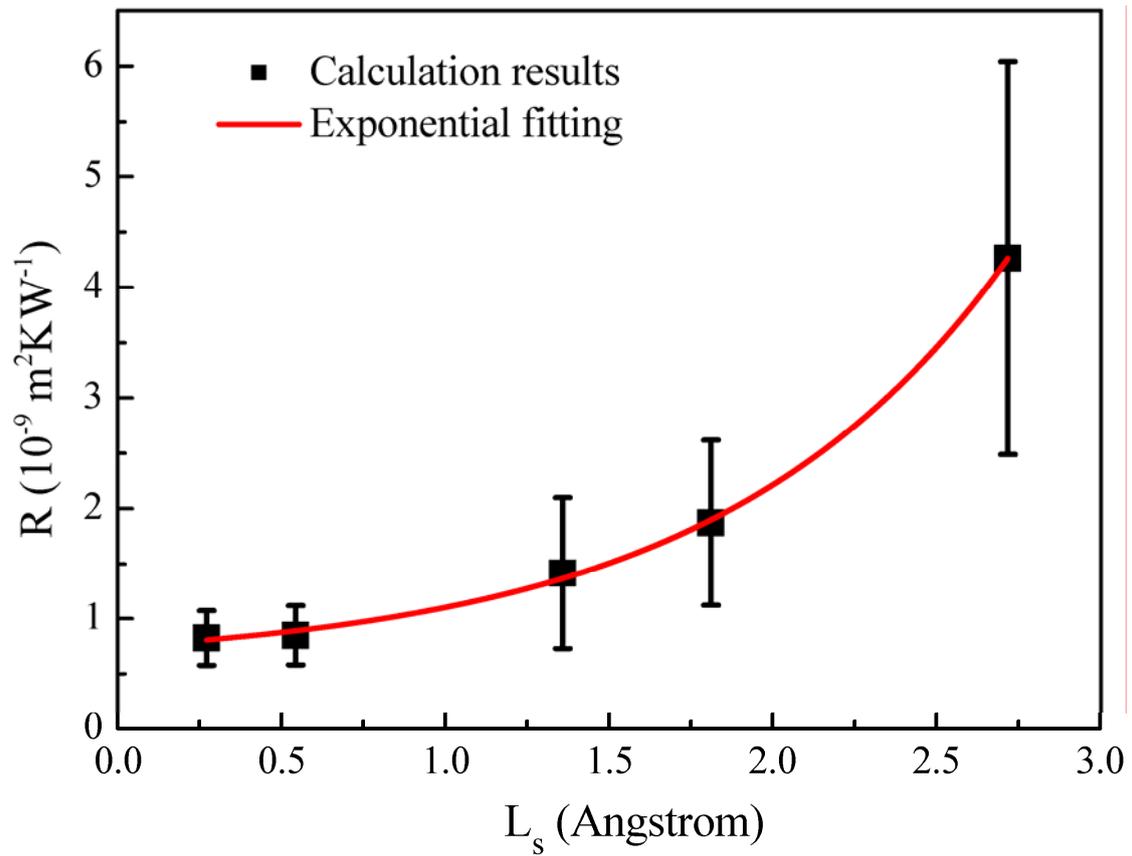

Figure 4. Kapitza resistance at $SiO_2$/Si interface versus separation distance between two segments at room temperature. The length of the heterojunction is about $L_x$=5.6 nm. The square and solid line denotes the calculation results and exponential fitting, respectively.

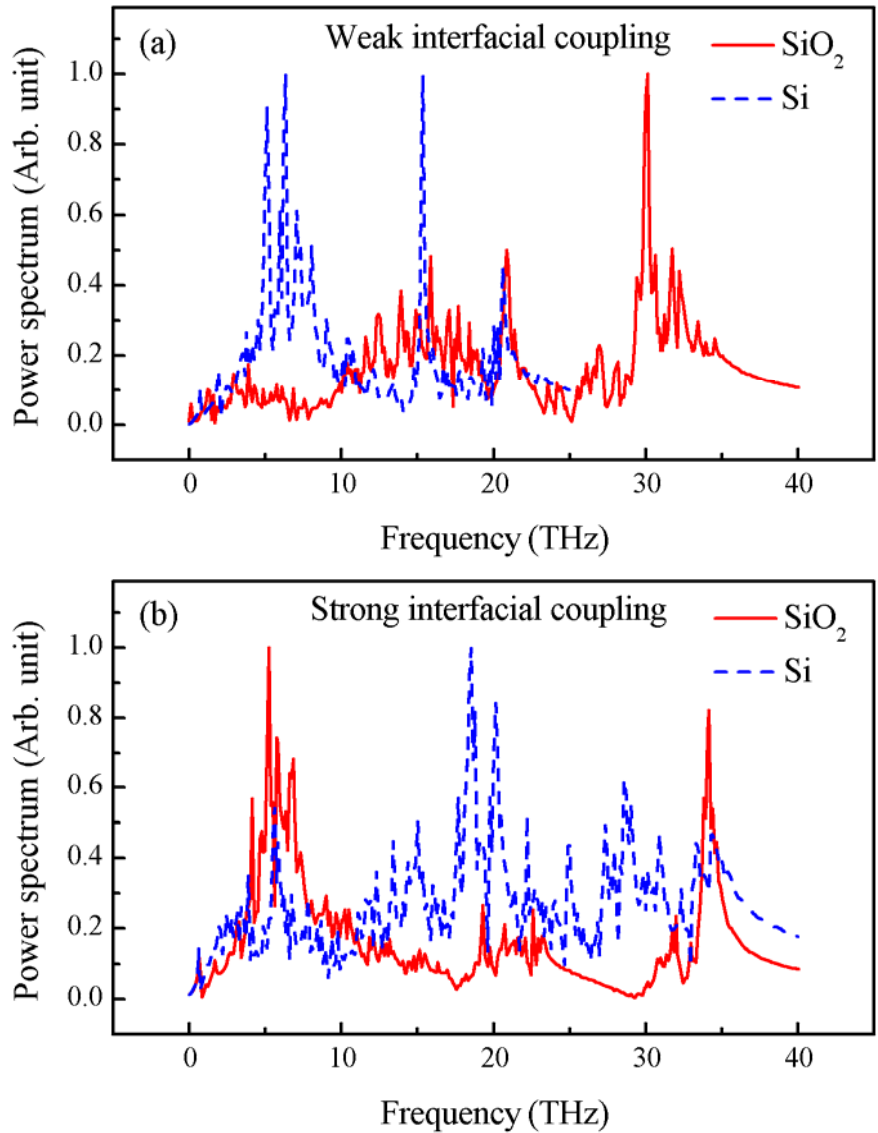

Figure 5. Typical power spectrum for atoms at each side of the SiO$_2$/Si interface with different interfacial coupling strength. The solid and dashed lines denote the power spectrum for SiO$_2$ and Si, respectively. (a) Weak interfacial coupling ($N_c$=2). (b) Strong interfacial coupling ($N_c$=20).

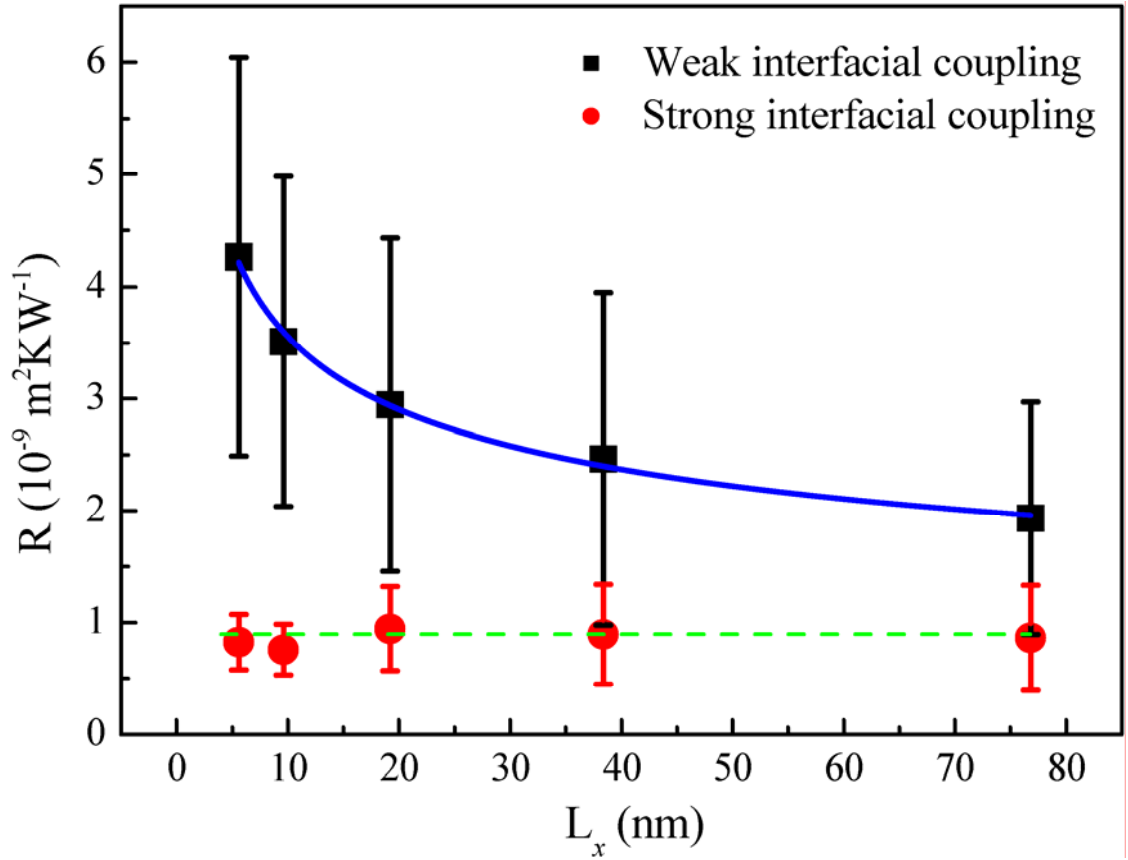

Figure 6. Kapitza resistance at $SiO_2$/Si interface versus the length of the heterojunction at room temperature. The square and circle denotes the calculation results for the weak ($N_c$=2) and strong ($N_c$=20) interfacial coupling case, respectively. The solid line draws the power law fitting for the weak interfacial coupling, and the dashed line draws the constant thermal resistance value of $0.9 \times 10^{-9}$ $m^2KW^{-1}$ for reference.

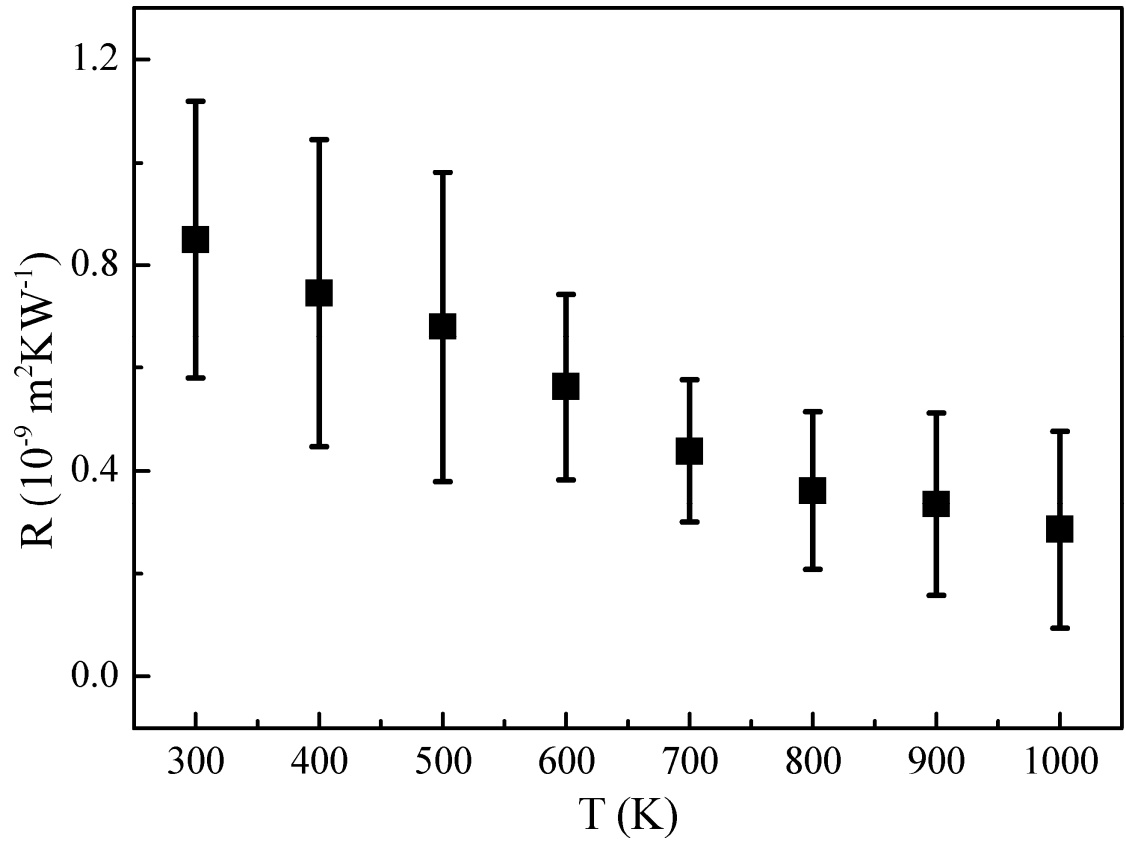

Figure 7. Temperature dependent Kapitza resistance at $SiO_2$/Si interface above room temperature for the strong interfacial coupling case ($N_c$=10). The length of the heterojunction is about $L_x$=5.6 nm.